\documentclass[conference]{IEEEtran}
\usepackage[latin1]{inputenc}
\usepackage{amsmath}
\usepackage{listings}
\usepackage{marvosym}
\usepackage{color}
\usepackage{graphicx}
\usepackage{subfigure}
\usepackage{multirow}
\usepackage{booktabs}
\usepackage{array}
\usepackage{bigstrut}
\usepackage{caption}
\usepackage{listings}
\usepackage{algorithm2e}
\usepackage{scalefnt}
\newtheorem{mydef}{Definition}

\newcommand{\comment}[1]{}

\usepackage{amssymb}
\usepackage{url}

\ifCLASSINFOpdf
\else
\fi
\begin{document}
\title{Model Checking Embedded C Software using \textit{k}-Induction and Invariants (extended version)}

% \author{
% \IEEEauthorblockN{Herbert Rocha}
% \IEEEauthorblockA{Federal University of Roraima\\
% Email: herbert.rocha@ufrr.br}
% \and
% \IEEEauthorblockN{Hussama Ismail}
% \IEEEauthorblockA{Federal University of Amazonas\\
% Email: hussamaismail@gmail.com}
% \and
% \IEEEauthorblockN{Lucas Cordeiro}
% \IEEEauthorblockA{Federal University of Amazonas\\
% Email: lucasccordeiro@gmail.com}
% \and
% \IEEEauthorblockN{Raimundo Barreto}
% \IEEEauthorblockA{Federal University of Amazonas\\
% Email: rbarreto@icomp.ufam.edu.br}
% }


\author{\IEEEauthorblockN{Herbert Rocha\IEEEauthorrefmark{1},
Hussama Ismail\IEEEauthorrefmark{2},
Lucas Cordeiro\IEEEauthorrefmark{2}, and
Raimundo Barreto\IEEEauthorrefmark{2}}
\IEEEauthorblockA{\IEEEauthorrefmark{1}Federal University of Roraima\\
Email: herbert.rocha@ufrr.br}
\IEEEauthorblockA{\IEEEauthorrefmark{2}Federal University of Amazonas\\
Email: \{hussamaismail,lucasccordeiro\}@gmail.com, rbarreto@icomp.ufam.edu.br}}
% \IEEEauthorblockA{\IEEEauthorrefmark{3}Federal University of Amazonas\\
% Email: lucasccordeiro@gmail.com}
% \IEEEauthorblockA{\IEEEauthorrefmark{4}Federal University of Amazonas\\
% Email: rbarreto@icomp.ufam.edu.br
% }}


\maketitle

\begin{abstract}
%\boldmath
We present a proof by induction algorithm, which combines \textit{k}-induction with invariants 
to model check embedded C software with bounded and unbounded loops. The \textit{k}-induction algorithm 
consists of three cases: in the base case, we aim to find a counterexample with up to \textit{k} loop unwindings; 
in the forward condition, we check whether loops have been fully unrolled and that the safety property $\phi$ holds 
in all states reachable within $k$ unwindings; and in the inductive step, we check that whenever $\phi$ holds 
for $k$ unwindings, it also holds after the next unwinding of the system. For each step of the \textit{k}-induction 
algorithm, we infer invariants using affine constraints (i.e., polyhedral) to specify pre- and post-conditions. 
Experimental results show that our approach can handle a wide variety of safety properties in 
typical embedded software applications from telecommunications, control systems, and medical devices; 
we demonstrate an improvement of the induction algorithm effectiveness if compared to other approaches.
\end{abstract}


% \fi
\IEEEpeerreviewmaketitle

% % % % % % % % % % % % % % % % % % % % % % % % % % % % % % % % % % % % % % % % % % % % %

% -----------------------------------------------------------------
% => INTRODUCTION -      1 -> [DOING]
% -----------------------------------------------------------------
% 
\vspace{-2ex}
%=-=-=-=-=-=-=-=-=-=-=-=-=-=-=-=-=-=-=-=-=-=-=-=-=-=-=
\section{Introduction}
%=-=-=-=-=-=-=-=-=-=-=-=-=-=-=-=-=-=-=-=-=-=-=-=-=-=-=

The Bounded Model Checking (BMC) techniques based on Boolean Satisfiability (SAT) 
% The Bounded Model Checking (BMC) techniques based on Boolean Satisfiability (SAT)~\cite{handbook09} 
or Satisfiability Modulo Theories (SMT)  
% or Satisfiability Modulo Theories (SMT)~\cite{BarrettSST09} 
have been applied to verify single- and multi-threaded programs and to find 
subtle bugs in real programs~\cite{CBMC2012,MerzFS12,Cordeiro12}. 
% subtle bugs in real programs~\cite{Clarke04,MerzFS12,TeseLucas}. 
The idea behind the BMC techniques is to check the negation
of a given property at a given depth, i.e., given a transition system \textit{M},
a property $ \phi $, and a limit of iterations \textit{k}, BMC unfolds the system
\textit{k} times and converts it into a Verification Condition (VC) $ \psi $ such that
$\psi$ is \textit{satisfiable} if and only if $\phi$ has a counterexample of depth
less than or equal to \textit{k}.

Typically, BMC techniques are only able to falsify properties up to 
a given depth \textit{k}; they are not able to prove the correctness 
of the system, unless an upper bound of \textit{k} is known, i.e., a bound that 
unfolds all loops and recursive functions to their maximum possible depth. 
In particular, BMC techniques limit the visited regions of data structures (e.g., arrays) 
and the number of loop iterations. This
limits the state space that needs to be explored during
verification, leaving enough that real errors in applications
\cite{CBMC2012,MerzFS12,Cordeiro12,Ivancic05} can be found; BMC tools are, however,
susceptible to exhaustion of time or memory limits for programs with loops
whose bounds are too large or cannot be determined statically.  

Consider for example the simple program in
Fig.~\ref{figure:unwindingprograms} (top), in which the loop in line~2 runs an
unknown number of times, depending on the initial
non-deterministic value assigned to \texttt{x} in line~1. The
assertion in line~3 holds independent of \texttt{x}'s initial value.
Unfortunately, BMC tools like CBMC \cite{CBMC2012}, LLBMC \cite{MerzFS12}, 
or ESBMC \cite{Cordeiro12} typically fail to verify programs that contain 
such loops. Soundness requires that they insert a so-called \emph{unwinding assertion} 
(the negated loop bound) at the end of the loop, as in 
Fig.~\ref{figure:unwindingprograms} (bottom), line $5$.
This \textit{unwinding assertion} causes the BMC tool to fail if $k$ is too small.

% %
% % \vspace{-3ex}
% \begin{figure}
% % [!htbp]
% \centering
% \begin{minipage}[t]{0.4\textwidth}
% \begin{lstlisting}
% unsigned int x=*;
% while(x>0) x--;
% assert(x==0);
% \end{lstlisting}
% \end{minipage}
% \hspace{1cm}
% \begin{minipage}[t]{0.4\textwidth}
% \begin{lstlisting}[escapechar=^]
% unsigned int x=*;
% if (x>0)
%   x--; 		^\raisebox{-1pt}[0pt][0pt]{$\Bigg\}\:k\mathrm{~copies}$}^
% ...     
% assert(!(x>0));
% assert(x==0);
% \end{lstlisting}
% \end{minipage}
% \caption{Unbounded loop (top) and finite unwinding (bottom)}
% \label{figure:unwindingprograms}
% \end{figure}
% % \vspace{-4ex}

In mathematics, one usually attacks such unbounded problems using \emph{proof by induction}.
A variant called \textit{k}-induction has been successfully 
combined with continuously-refined invariants~\cite{Beyer:2015arxiv}, to prove that (restricted) 
C programs do not contain data races~\cite{Donaldson10,Kinductor}, or that
design-time time constraints are respected~\cite{EenS03}.
Additionally, \textit{k}-induction is a well-established technique in hardware verification,
where it is easy to apply due to the monolithic transition relation present
in hardware designs~\cite{EenS03,GrosseLD09,Sheera00}. This paper
contributes a new algorithm to prove correctness 
of C programs by \textit{k}-induction in a completely 
automatic way.

\vspace{-2ex}
\begin{figure}[!htbp]
\centering
\begin{minipage}[t]{0.4\textwidth}
\begin{lstlisting}
unsigned int x=*;
while(x>0) x--;
assert(x==0);
\end{lstlisting}
\end{minipage}
\hspace{1cm}
\begin{minipage}[t]{0.4\textwidth}
\begin{lstlisting}[escapechar=^]
unsigned int x=*;
if (x>0)
  x--; 		^\raisebox{-1pt}[0pt][0pt]{$\Bigg\}\:k\mathrm{~copies}$}^
...     
assert(!(x>0));
assert(x==0);
\end{lstlisting}
\end{minipage}
\caption{Unbounded loop (top) and finite unwinding (bottom)}
\label{figure:unwindingprograms}
\end{figure}
\vspace{-2ex}

The main idea of the algorithm is to use an iterative deepening approach and check,
for each step \textit{k} up to a maximum value, three different cases called here as base case, forward condition,
and inductive step. Intuitively, in the base case, we intend to find a counterexample of $\phi$ with up to
\textit{k} iterations of the loop. The forward condition checks whether loops have been fully unrolled
and the validity of the property $\phi$ in all states reachable within \textit{k} iterations. 
The inductive step verifies that if $\phi$ is valid for
\textit {k} iterations, then $\phi$ will also be valid for the next unfolding of the system. 
For each step, we infer invariants using affine constraints
to prune the state space exploration and to strengthen the induction hypothesis.

These algorithms were all implemented in the Efficient SMT-based
Context-Bounded Model Checker tool (known as ESBMC), 
which uses BMC techniques and SMT solvers to verify
embedded systems written in C/C$++$~\cite{Cordeiro12,Ramalho2013}.
In Cordeiro et al.~\cite{Cordeiro12,Ramalho2013} the ESBMC tool is presented, which describes
how the input program is encoded in SMT; 
what the strategies for unrolling loops are; what are the transformations/optimizations that are
important for performance; what are the benefits of using an SMT solver
instead of a SAT solver; and how counterexamples to falsify properties are
reconstructed. Here we extend our previous work and focus our contribution 
on the combination of the \textit{k}-induction algorithm with invariants.  
First, 
we describe the details of an accurate translation that extends ESBMC 
to prove the correctness of a given (safety) property for any depth without manual
annotations of loops invariants.
%
% Second, we adopt program invariants (using polyhedra) in the \textit{k}-induction algorithm, 
% to speed up the verification time and to improve the quality 
% of the results by solving more verification tasks in less time.
% 
Second, we adopt program invariants (using polyhedra) in the \textit{k}-induction algorithm, 
to improve the quality of the results by solving more verification tasks. 
% 
% we use a multi-process implementation
% of the \textit{k}-induction algorithm, similar to Kahsai et al.~\cite{Kahsai11}, 
% to speedup the verification time and to improve the quality 
% of the results by solving more verification tasks in less time. 
%
Third, we show that our implementation is applicable to a broader range of verification tasks;
in particular embedded systems, where existing approaches do not 
to support~\cite{Donaldson10,Kinductor,GrosseLD09}.

% To validate the implementations of the algorithm, we used the loops \allowbreak benchmarks from the
% International Competition on Software Verification (SV-COMP)~\cite{svcomp_book:2013}.
% 
% The experimental results show that, both the implementations with and without invariants 
% using polyhedral, are able to verify a wide variety of safety properties. 
% The experiments also show that the implementation with invariants presents better results,
% which is able to prove and falsify more properties in the verification tasks, 
% while it requires less verification time.

% -----------------------------------------------------------------
% => BACKGROUND        - 2 -> [DOING]
% -----------------------------------------------------------------
% 
%=-=-=-=-=-=-=-=-=-=-=-=-=-=-=-=-=-=-=-=-=-=-=-=-=-=-=
\section{Induction-based Verification of C Programs using Invariants}
\label{sec:kinduction}
%=-=-=-=-=-=-=-=-=-=-=-=-=-=-=-=-=-=-=-=-=-=-=-=-=-=-=

The transformations in each step of the \textit{k}-induction algorithm 
take place at the intermediate representation level, after 
converting the C program into a GOTO-program, which simplifies the representation
and handles the unrolling of the loops and the elimination of recursive functions.

%=-=-=-=-=-=-=-=-=-=-=-=-=-=-=-=-=-=-=-=-=-=-=-=-=-=-=
\subsection{The Proposed \textit{k}-Induction Algorithm}
\label{sec:k-induction-algorithm}
%=-=-=-=-=-=-=-=-=-=-=-=-=-=-=-=-=-=-=-=-=-=-=-=-=-=-=

Figure~\ref{figure:k-induction-algorithm} shows an overview of the proposed \textit{k}-induction 
algorithm. The input of the algorithm is a C program $P$
together with the safety property $\phi$.~The algorithm returns \textit{TRUE} 
(if there is no path that violates the safety property), \textit{FALSE} (if there exists a path that 
violates the safety property), and \textit{UNKNOWN} (if it does not succeed in computing an answer 
\textit{true} or \textit{false}).

In the base case, the algorithm tries to find a counterexample up to a 
maximum number of iterations \textit{k}.~In the forward 
condition, global correctness of the loop w.r.t. the property 
is shown for the case that the loop iterates at most $k$ times;
and in the inductive step, the algorithm checks that, if the property is valid in \textit{k} iterations,
then it must be valid for the next iterations. The algorithm runs up to a maximum number of iterations
and increases the value of $k$ if it cannot falsify the property. 
 In Figure~\ref{figure:k-induction-algorithm}, the algorithm also performs a rechecking/refinement of the result 
(using the flag \texttt{force\_basecase}) by the BMC procedure. In particular, we re-check the results 
in the forward condition and the inductive step (adopting an increment of the actual $k$). 
This re-checking procedure is needed due to the inclusion of invariants, 
which over-approximates the analyzed program; otherwise, the invariants could result in 
incorrect exploration of the states sets. 
\begin{figure}
\centering
\begin{minipage}{0.4\textwidth}
\begin{lstlisting} [escapechar=^]
input: program P and safety property ^$\phi$^
output: true, false, or unknown
k = 1
force_basecase = FALSE 
last_result = UNKNOWN
while k <= max_iterations do
  if force_basecase then
     k = k + 5  
  if base_case(P, ^$\phi$^, k) then
	  show counterexample s[0..k]
    return FALSE
  else
    if force_basecase then 
       return last_result
    k=k+1    
    if forward_condition(P, ^$\phi$^, k) then
       force_basecase = TRUE
       last_result = TRUE
    else
      if inductive_step(P, ^$\phi$^, k) then
         force_basecase = TRUE 
         last_result = TRUE
      end-if
    end-if
  end-if
end-while
return UNKNOWN
\end{lstlisting}
\end{minipage}
\caption{An overview of the \textit{k}-induction algorithm.}
\label{figure:k-induction-algorithm}
\end{figure}
% \vspace{-2ex}

%=-=-=-=-=-=-=-=-=-=-=-=-=-=-=-=-=-=-=-=-=-=-=-=-=-=-=
\subsubsection{Loop-free Programs}
\label{sec:multiple-nested-loops}
%=-=-=-=-=-=-=-=-=-=-=-=-=-=-=-=-=-=-=-=-=-=-=-=-=-=-=

In the \textit{k}-induction algorithm, the loop 
unwinding of the program is done incrementally from one 
to \textit{max\_iterations} (cf. Fig.~\ref{figure:k-induction-algorithm}), 
where the number of unwindings is 
% measured by counting the number of \textit{backjumps}~\cite{Muchnick:1998}.
measured by counting the number of \textit{backjumps}~\cite{Muchnick:1998}. 
In each step, an instance 
of the program that contains $k$ copies 
of the loop body corresponds to checking a loop-free program, 
which uses only \textit{if}-statements in order to prevent its execution 
in the case that the loop ends before $k$ iterations. 

\begin{mydef}
\label{loop-free-program}
\textbf{(Loop-free Program)}\textit{ A loop-free program is represented by a straight-line program 
(without loops) by providing an $ite\left(\theta, \rho_1, \rho_2\right)$ 
operator, which takes a Boolean formula $\theta$ and, depending on its value, 
selects either the second $\rho_1$ or the third argument $\rho_2$,
where $\rho_1$ represents the loop body and $\rho_2$ represents either another
$ite$ operator, which encodes a \textit{k}-copy of the loop body, or an assertion/assume
statement.}
\end{mydef}

Each step of our \textit{k}-induction algorithm (except for the base case) 
transforms a program with loops into a loop-free program, such that the correctness 
of the loop-free program implies the correctness of the program with loops.

If the program consists of multiple 
and possibly nested loops, we simply set the number of loop unwindings 
globally, that is, for all loops in the program and apply these aforementioned 
translations recursively. 
% 
% Figure~\ref{figure:iteration-based-unwinding} shows 
% how loop unwindings are applied to a program with nested loops. 
% 
Note that each case of the \textit{k}-induction algorithm performs different transformations 
at the end of the loop: either to find bugs (base case) or to prove that enough 
loop unwindings have been done (forward condition).
%
% \begin{figure*}
% \centering
% \includegraphics[scale=0.45]{figuras/iteration-based-unwinding.jpg}
% \caption{(a) A program with nested loops. (b) Iteration-based unwinding of the program in (a).}
% \label{figure:iteration-based-unwinding}
% \end{figure*}

%=-=-=-=-=-=-=-=-=-=-=-=-=-=-=-=-=-=-=-=-=-=-=-=-=-=-=
\subsubsection{Program Transformations}
\label{sec:program-translations}
%=-=-=-=-=-=-=-=-=-=-=-=-=-=-=-=-=-=-=-=-=-=-=-=-=-=-=

In terms of program transformations, which are all done completely automatically 
by our proposed method, the base case simply inserts 
an unwinding assumption, to the respective loop-free program $P'$, consisting 
of the termination condition $\sigma$ after the loop, as follows $I \wedge T \wedge  \sigma \Rightarrow \phi \nonumber$, 
where $I$ is the initial condition, $T$ is the transition
relation of $P'$, and $\phi$ is a safety
property. 

The forward case inserts an unwinding assertion 
instead of an assumption after the loop, as follows $I \wedge T \Rightarrow \sigma \wedge \phi \nonumber$.
Our base case and forward condition translations are implemented 
on top of plain BMC. 
However, for the inductive step of the algorithm, 
several transformations are carried out. In particular, the loop
$while(c) \left\{ E; \right\}$ is converted into
\begin{equation}
\label{eqwhile-code-transformed}
  \begin{array}{c}
    A; while(c) \left\{ S; E; U; \right\} R;
  \end{array}
\end{equation}
\noindent where $A$ is the code responsible for 
assigning non-deterministic values to all loop variables,
i.e., the state is havocked before the loop, $c$ is the exit condition of the loop \textit{while}, 
$S$ is the code to store the current state of the program variables 
before executing the statements of $E$, $E$ is the actual code inside the 
loop \textit{while}, $U$ is the code to update all program variables
with local values after executing $E$, and $R$ is the code to remove redundant states.
\begin{mydef}
\label{loop-variable}
\textbf{(Loop Variable)}\textit{ A loop variable is a variable $v \subseteq V$, where 
$V = V_{global}\cup V_{local}$ given that $V_{global}$ is the set of global variables 
and $V_{local}$ is the set of local variables that occur in the loop of a program.}
\end{mydef}
\begin{mydef}
\label{havoc-loop-variable}
\textbf{(Havoc Loop Variable)}\textit{ A nondeterministic value is assigned to a 
loop variable $v$ if and only if $v$ is used in the loop termination condition 
$\sigma$, in the loop counter that controls iterations of a loop, 
or modified inside the loop body.}
\end{mydef}
The intuitive interpretation of $S$, $U$, and $R$ is that if the current state (after executing $E$)
is different than the previous state (before executing $E$), then new states are produced in the
given loop iteration; otherwise, they are redundant and the code $R$ is then 
responsible for preventing those redundant states to be included into the states vector. 
Note further that the code $A$ assigns non-deterministic values
to all loop variables so that the model checker can explore 
all possible states implicitly. 
Similarly, the loop \textit{for} can easily be converted into the loop \textit{while} as follows:
$for(B; c; D) \left\{ E; \right\}$ is rewritten as
\begin{equation}
\label{eqfor-code-transformedt}
  \begin{array}{c}
    B; \: while(c) \left\{ E; D; \right\}
  \end{array}
\end{equation}
\noindent where $B$ is the initial condition of the loop, $c$ is the exit condition of the loop,
$D$ is the increment of each iteration over $B$, and $E$ is the actual code inside the loop \textit{for}. 
No further transformations are applied to the loop \textit{for} during the inductive step.
Additionally, the loop \textit {do while} can be 
converted into the loop \textit{while} with one difference, 
the code inside the loop must execute
at least once before the exit condition is checked.

The inductive step is thus represented by
$\gamma \wedge \sigma \Rightarrow \phi$,
where $\gamma$ is the transition relation of $\hat{P'}$, which represents 
a loop-free program (cf. Definition~\ref{loop-free-program}) after applying 
transformations (\ref{eqwhile-code-transformed}) and (\ref{eqfor-code-transformedt}).
The intuitive interpretation of the inductive step is to prove that,
for any unfolding of the program, there is no assignment of particular 
values to the program variables that violates the safety property being checked.
Finally, the induction hypothesis of the inductive step consists of the 
conjunction between the postconditions ($Post$) and the 
termination condition ($\sigma$) of the loop.

%=-=-=-=-=-=-=-=-=-=-=-=-=-=-=-=-=-=-=-=-=-=-=-=-=-=-=
\subsubsection{Invariant Generation}
\label{sec:invariant-gen}
%=-=-=-=-=-=-=-=-=-=-=-=-=-=-=-=-=-=-=-=-=-=-=-=-=-=-=

% To infer program invariants, we adopted the PIPS~\cite{Maisonneuve:201417,pips:2013} tool, which
To infer program invariants, we adopted the PIPS~\cite{pips:2013} tool, which 
is an interprocedural source-to-source compiler framework for C and Fortran programs and 
relies on a polyhedral abstraction of program behavior. 
PIPS performs a two-step analysis: (1) each program instruction is associated to an affine transformer, 
representing its underlying transfer function. This is a bottom-up procedure, starting from elementary instructions, 
then working on compound statements and up to function definitions; (2) polyhedral invariants are propagated along 
with instructions, using previously computed transformers.

In our proposed method, PIPS receives the analyzed program as input and then it generates 
invariants that are given as comments surrounding instructions in the output C code. 
These invariants are translated and instrumented into the program as assume statements. 
In particular, we adopt the function \texttt{assume($expr$)} 
to limit possible values of the variables that are related to the invariants.
This step is needed since PIPS generates invariants that are presented as mathematical expressions 
(e.g., $2j < 5t$), which are not accepted by C programs syntax and invariants with $\#init$ 
suffix that is used to distinguish the old value from the new value.

% [DOING] Talk how is handled the PIPS output?
Algorithm~\ref{alg:tradinvpips} shows the proposed method, which receives 
as inputs the code generated by PIPS (\texttt{PIPSCode}) with invariants as comments, 
and it generates as output a new instance of the analyzed code 
(\texttt{NewCodeInv}) with invariants, adopting the function \texttt{assume($expr$)}, where $expr$ 
is an expression supported by the C programming language. 
The time complexity of this algorithm is $O(n^2)$, where $n$ 
is code size with invariants generated by PIPS. 
The algorithm is split into three parts: 
(1) identify the \texttt{\#init} structure in the PIPS invariants; 
(2) generate code to support the translation of the \texttt{\#init} structure in the 
the invariant; and finally
(3) translate mathematical expressions contained in the invariants, which is 
performed by the invariants transformation in the PIPS format to the C programming language. 

\IncMargin{1em}
\RestyleAlgo{boxed}
\LinesNumbered

\begin{algorithm}[!htb]
\scalefont{0.8}
\caption{Translation algorithm of invariants} 
\label{alg:tradinvpips}
\SetAlgoLined

\KwIn{PIPSCode - C code with PIPS invariants}
\KwOut{NewCodeInv - New code with invariant supported by C programs}                    

% \tcp{dictionary to identify \#init}
dict\_variniteloc $\leftarrow$ \{ \}

% \tcp{list for the new code generated in the translation}
NewCodeInv $\leftarrow$ \{ \}

\tcp{Part 1 - identifying \#init}
\ForEach{line of the PIPSCode}{ \label{Line:readCodeIdInitpt1}

%   Identify #init from PIPS in the code with invariants
  \If{is a PIPS comment in this pattern \texttt{// P(w,x) \{$w==0$, $x\#init>10$\}}} { \label{Line:pt1_invpips}
    \If{the comment has the pattern \texttt{([a-zA-Z0-9\_]+)\#init} } {
      dict\_variniteloc[line] $\leftarrow$ the variable suffixed \#init \label{Line:pt1_idint}
    }    
  }
  
}

\tcp{Part 2 - code generation}
\ForEach{line of PIPSCode}{ \label{Line:readCodeWriInitpt2}

  NewCodeInv $\leftarrow$ line
  
  \If{is the beginning of a function} {    
    \If{has some line number of this function $\in$ dict\_variniteloc}{ \label{Line:pt2_funcwithinit}
      \ForEach{variable $\in$ dict\_variniteloc}{
	NewCodeInv $\leftarrow$ Declare a variable with this pattern \texttt{type var\_init $=$ var\;}
      }
    }    
  }
  
}

\tcp{Part 3 - correct the invariant format}
\ForEach{line of NewCodeInv}{ \label{Line:readCodeGenInvpt3}
  
%   \tcp{list to the translated invariants}
  listinvpips $\leftarrow$ \{ \}
  
  NewCodeInv $\leftarrow$ line
    
  \If{is a PIPS comment in this pattern \texttt{// P(w,x) \{$w==0$, $x\#init>10$\}}} {
    \ForEach{expression $\in$ \texttt{\{$w==0$, $x\#init>10$\}}}{
      listinvpips $\leftarrow$ Reformulate the expression according to the C programs syntax and 
      replace $\#init$ by $\_init$ \label{Line:pt3_regextrans}
    }
    
    NewCodeInv $\leftarrow$ \texttt{\_\_ESBMC\_assume}(concatenate the invariants in  
    listinvpips with $\&\&$)
  }
    
}

\end{algorithm}

Line~\ref{Line:readCodeIdInitpt1} of Algorithm~\ref{alg:tradinvpips} 
performs the first part of the invariant translation, which consists of reading each line 
of the analyzed code with invariants and identifying whether a given comment is an invariant 
generated by PIPS (line~\ref{Line:pt1_invpips}). If an invariant is identified and it contains 
the structure \texttt{\#int}, then the invariant location (the line number) is stored, 
as well as, the type and name of the variable, which has the prefix 
\texttt{\#int} (line~\ref{Line:pt1_idint}).

After identifying the \texttt{\#int} structures in the invariants, 
the second part of Algorithm~\ref{alg:tradinvpips} performs 
line~\ref{Line:readCodeWriInitpt2}, which consists of reading each line of the 
analyzed code with invariants (\texttt{PIPSCode}), and identifying the beginning 
of each function in the code. For each identified function, the algorithm checks whether 
that function has identified some \texttt{\#int} structure  (line~\ref{Line:pt2_funcwithinit}). 
If it has been identified, for each variable that has the suffix \texttt{\#int}, a new line of 
code is generated at the beginning of the function, with the declaration of an auxiliary variable, 
which contains the old variable value, i.e., its value at the beginning of the function. 
% The new created variable has the following format \texttt{type var\_init $=$ var\;}, where 
% \texttt{type} is the identified variable type, and \texttt{var} is the identified variable name. 
During the execution of this algorithm, a new instance of the code (\texttt{NewCodeInv}) is generated.

In the third (and final part) of Algorithm~\ref{alg:tradinvpips} (line~\ref{Line:readCodeGenInvpt3}), 
each line of the new code instance (\texttt{NewCodeInv}) is read to convert
each PIPS invariant into expressions supported by the C language. 
This transformation consists in applying regular expressions (line~\ref{Line:pt3_regextrans}) to  
add operators (e.g., from $2j$ to $2*j$) and replacing the structure \texttt{\#int} to \texttt{\_int}. 
For each analyzed PIPS comment/invariant, we generate a new code line to the new format, where this line 
is concatenated with the operator \texttt{\&\&} and added to the \texttt{\_\_ESBMC\_assume} function.

\section{Experimental Evaluation}
\label{sec:exp}
%=-=-=-=-=-=-=-=-=-=-=-=-=-=-=-=-=-=-=-=-=-=-=-=-=-=-=

This section is split into two parts.
The setup is described in Section~\ref{experimental-setup}
and Section~\ref{results-kinduction} describes a comparison
among DepthK~\footnote{https://github.com/hbgit/depthk}, ESBMC~\cite{Cordeiro12}, CBMC~\cite{CBMC2012}, and CPAchecker~\cite{Beyer:2015arxiv} using a set of C benchmarks from SV-COMP~\cite{svcomp2015} and embedded applications~\cite{WCET:2012,Scott:1998powerstone,SNU:2012}.

%--------------------------------------
\subsection{Experimental Setup}
\label{experimental-setup}
%--------------------------------------

The experimental evaluation is conducted on a computer with Intel Xeon CPU $E5-2670$ CPU, 
$2.60$GHz, $115$GB RAM with Linux $3.13.0-35$-generic x$86$\_$64$. 
Each verification task is limited to a CPU time of $15$ minutes and a 
memory consumption of $15$ GB. Additionally, we defined the \textit{max\_iterations} 
to $100$ (cf. Fig.~\ref{figure:k-induction-algorithm}). 
To evaluate all tools, we initially adopted $142$ ANSI-C programs of the SV-COMP $2015$ benchmarks; 
in particular, the \textit{Loops} subcategory; and 
$34$ ANSI-C programs used in embedded systems: 
Powerstone~\cite{Scott:1998powerstone} contains a set of C programs for embedded systems (e.g., 
for automobile control and fax applications); while SNU real-time~\cite{SNU:2012} contains
a set of C programs for matrix and signal processing functions such as matrix
multiplication and decomposition, quadratic equations
solving, cyclic redundancy check, fast fourier transform,
LMS adaptive signal enhancement, and JPEG encoding;
and the WCET~\cite{WCET:2012} contains C programs adopted for 
worst-case execution time analysis.

We also present a comparison with the tools: 
DepthK v$1.0$ with $k$-induction and invariants using polyhedra, the parameters are 
defined in the wrapper script available in the tool repository; 
ESBMC v$1.25.2$ adopting $k$-induction without invariants. We adopted the wrapper script from 
SV-COMP 2013\footnote{http://sv-comp.sosy-lab.org/2013/} to execute the tool; 
CBMC v$5.0$ with $k$-induction, running the script provided in \cite{Beyer:2015arxiv}; 
CPAChecker\footnote{ https://svn.sosy-lab.org/software/cpachecker/trunk} with $k$-induction and invariants 
at revision $15596$ from its SVN repository. The options to execute the tool are defined 
in \cite{Beyer:2015arxiv}. 
%To improve the presentation, we report only the results of the options that 
%presented the best results. These options are defined in \cite{CPAchecker_boostinv:2015} as follows: 
%CPAchecker \textit{cont.-ref. k-Induction} (\textit{k-Ind InvGen}) and CPAchecker \textit{no-inv k-Induction}. 

%=-=-=-=-=-=-=-=-=-=-=-=-=-=-=-=-=-=-=-=-=-=-=-=-=-=-=
\subsection{Experimental Results}
\label{results-kinduction}
%=-=-=-=-=-=-=-=-=-=-=-=-=-=-=-=-=-=-=-=-=-=-=-=-=-=-=
\vspace{-10ex}
\begin{center}
\begin{table*}
\centering 
\newcolumntype{M}[1]{>{\centering\let\newline\\\arraybackslash\hspace{0pt}}m{#1}}
\begin{tabular}{|l|M{1.5cm}|M{1.5cm}|M{2cm}|M{3cm}|M{1.5cm}|}
\hline
\textbf{Tool} & DepthK  & ESBMC + k-induction & CPAchecker no-inv k-Induction & CPAchecker cont.-ref. k-Induction (k-Ind InvGen) & CBMC + k-induction \bigstrut\\
\hline
\textbf{Correct Results} & 94     & 70    & 78    & 76    & 64 \bigstrut\\
\hline
\textbf{False Incorrect} & 1     & 0     & 0     & 1     & 3 \bigstrut\\
\hline
\textbf{True Incorrect} & 0     & 0     & 4     & 7     & 1 \bigstrut\\
\hline
\textbf{Unknown and TO} & 47     & 72    & 60    & 58    & 74 \bigstrut\\
\hline
\textbf{Time} & 190.38min  & 141.58min & 742.58min & 756.01min & 1141.17min \bigstrut\\
\hline
\end{tabular} 
\caption{Experimental results for the SVCOMP'15 loops subcategory.}
\label{Table:resultSVCOMPLoops}
\end{table*}
\end{center}
\begin{center}
\begin{table*}
\centering
\newcolumntype{M}[1]{>{\centering\let\newline\\\arraybackslash\hspace{0pt}}m{#1}}
\begin{tabular}{|l|M{1.5cm}|M{1.5cm}|M{2cm}|M{3cm}|M{1.5cm}|}
\hline
\textbf{Tools} & DepthK & ESBMC + k-induction & CPAchecker no-inv k-Induction & CPAchecker cont.-ref. k-Induction (k-Ind InvGen) & CBMC + k-induction \bigstrut\\
\hline
\textbf{Correct Results} & 17     & 18    & 27    & 27    & 15 \bigstrut\\
\hline
\textbf{False Incorrect} & 0     & 0     & 0     & 0     & 0 \bigstrut\\
\hline
\textbf{True Incorrect} & 0     & 0     & 0     & 0     & 0 \bigstrut\\
\hline
\textbf{Unknown and TO} & 17     & 16    & 7     & 7     & 19 \bigstrut\\
\hline
\textbf{Time} & 77.68min  & 54.18min & 1.8min & 1.95min & 286.06min \bigstrut\\
\hline
\end{tabular}
\caption{Experimental results for the Powerstone, SNU, and WCET benchmarks.}
\label{Table:resultEmbedded}
\end{table*}
\end{center}

After running all tools, we obtained the results shown in 
Table~\ref{Table:resultSVCOMPLoops} for the SV-COMP $2015$ benchmark
and in Table~\ref{Table:resultEmbedded} for the embedded systems benchmarks, 
where each row of these tables means: 
name of the tool (Tool); 
total number of programs that satisfy the specification (correctly) identified by the tool (Correct Results); 
total number of programs that the tool has identified an error for a program that meets the 
specification, i.e., false alarm or incomplete analysis (False Incorrect); 
total number of programs that the tool does not identify an error, i.e., 
bug missing or weak analysis (True Incorrect); 
Total number of programs that the tool is unable to model check due to lack of 
resources, tool failure (crash), or the tool exceeded the verification time of $15$ min (Unknown and TO); 
the run time in minutes to verify all programs (Time). 

We evaluated all tools as follows: for each program we identified the verification result 
and time. We adopted the same scoring 
scheme that is used in SVCOMP $2015$\footnote{http://sv-comp.sosy-lab.org/2015/rules.php}. 
For every bug found, $1$ score is assigned, for every correct safety proof, $2$ 
scores are assigned. A score of $6$ is subtracted for every wrong alarm (False Incorrect), 
and $12$ scores are subtracted for every wrong safety proof 
(True Incorrect). According to \cite{Beyer:2015arxiv}, this scoring scheme gives much more 
value in proving properties than finding counterexamples, and significantly punishes wrong answers
to give credibility for tool developers. 
It is noteworthy that for the embedded systems programs, we have used safe programs~\cite{Cordeiro12}
since we intend to check whether we produce strong invariants to prove properties.

The experimental results related to \textit{Loops} benchmarks had shown 
that the best scores belong to the DepthK, which combines $k$-induction 
with invariants, achieving  $140$ scores, ESBMC with $k$-induction without 
invariants achieved $105$ scores, CPAchecker \textit{no-inv} $k$-\textit{induction} 
achieved $101$ scores, and CBMC achieved $53$ scores. 
In the embedded systems benchmarks, we found that the best scores belong to 
the CPAchecker \textit{no-inv} $k$-\textit{induction} with $54$ scores, ESBMC 
with $k$-induction without invariants achieved $36$ scores,  
DepthK combined with $k$-induction and invariants, achieved $34$ scores, and 
CBMC achieved $30$ scores. 

We observed that DepthK achieved a lower score in the 
embedded systems benchmarks. However, the DepthK results are still higher than that of CBMC; 
and in the SV-COMP benchmark, DepthK achieved the highest score among all tools. 
In DepthK, we identified that, in turns, the low score in the 
embedded systems benchmarks is due to $35.30$\% of the results identified as \texttt{Unknown}, 
i.e., when it is not possible to determine an outcome or due to a tool failure. 
We also identified failures related to invariant generation 
and code generation that is given as input to the BMC procedure. 
It is noteworthy that DepthK is still under development (in a preliminary state), so we argue that the results are promising.

To measure the impact of applying invariants to the $k$-induction based verification, 
we classified the distribution of the DepthK and ESBMC results, per verification step, 
i.e., base case, forward condition, and inductive step. 
Additionally, we included the verification tasks that result in \texttt{unknown} 
and \texttt{timeout}. 
In this analysis, we evaluate only the results of DepthK and ESBMC, because they 
are part of our solution, and also because in the other tools, it is not possible to identify 
the steps of the $k$-induction in the standard logs generated by each tool. 

The result distribution shows that DepthK can prove more than $25.35$\% and $29.41$\% of the loops 
and embedded systems properties than ESBMC during the inductive step, respectively. 
These results lead to the conclusion that invariants help the \textit{k}-induction algorithm to prove 
more properties. We also identified that DepthK did not find a solution in $33.09$\% of the programs in 
the SV-COMP benchmarks and $50$\% in the embedded systems benchmarks (producing \texttt{Unknown} and \texttt{Timeout}).
This is explained due to the invariants generated from PIPS, which are not strong enough
for the verification with the $k$-induction, either due to a transformer or due to the invariants that 
are not convex; and also due to some errors in the tool implementation. 
ESBMC with $k$-induction did not find a solution in $50.7$\% 
of the programs in 
% Figure~\ref{figure:check_kstep_loops}, 
the SV-COMP benchmark, 
i.e., $17.61$\% more than DepthK (adding \texttt{Unknown} and \texttt{Timeout}); 
and in 
% Figure~\ref{figure:check_kstep_emebedded}, 
the embedded benchmarks, ESBMC did not find a solution 
in $47.06$\%, then only $3.64$\% less than the DepthK, thus providing evidences 
that the program invariants combined with $k$-induction can improve the verification results.

\begin{figure*}[!htbp]
      \centering
      \includegraphics[scale=0.4]{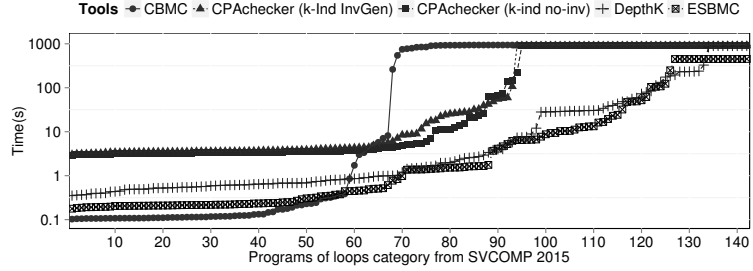}
      \caption{Verification time to the loops subcategory.}
      \label{figure:time_loops}
\end{figure*}

\begin{figure*}[!htbp]
      \centering
      \includegraphics[scale=0.4]{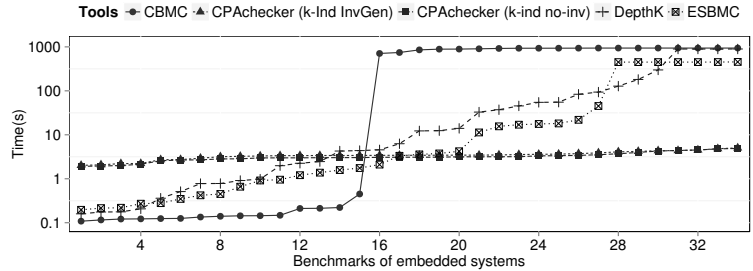}
      \caption{Verification time to the embedded systems programs.}
      \label{figure:time_embedded}
\end{figure*}

In Table~\ref{Table:resultSVCOMPLoops}, the verification time of DepthK to the loops benchmarks is usually faster than the other tools, except for ESBMC, as shown in Figure~\ref{figure:time_loops}. This happens because DepthK has an additional time for the invariants generation. 
In Table~\ref{Table:resultEmbedded}, we identified that the verification time of DepthK is only faster than CBMC (see Figure~\ref{figure:time_embedded}). However, note that the DepthK verification time is proportional to ESBMC, since the time difference is $23.5$min; we can argue that this difference is associated to the DepthK invariant generation. 

We believe that DepthK verification time can be improved in two directions:
fix errors in the tool implementation, because some results generated as 
\texttt{Unknown} are related to failures in the tool execution; and adjustments 
in the PIPS script parameters to generate stronger invariants, since PIPS has a broad 
set of commands for code transformation, parameter tuning might have a positive impact. 

% -----------------------------------------------------------------
% RELATED WORK         - 4 -> [DOING]
% -----------------------------------------------------------------
% 
% \newpage
%=-=-=-=-=-=-=-=-=-=-=-=-=-=-=-=-=-=-=-=-=-=-=-=-=-=-=
\section{Related Work}
\label{sec:related}
%=-=-=-=-=-=-=-=-=-=-=-=-=-=-=-=-=-=-=-=-=-=-=-=-=-=-=

The \textit{k}-induction application is gaining popularity in the software verification community. Recently, Bradley et al. introduce ``property-based reachability'' (or IC$3$) procedure for the safety verification of systems~\cite{Bradley12,Bradley13}. The authors have shown that IC$3$ can scale on certain benchmarks, where \textit{k}-induction fails to succeed. However, we do not compare \textit{k}-induction against
IC$3$ since it is already done by Bradley~\cite{Bradley12}; we focus our comparison on related \textit{k}-induction procedures.

Previous work on the one hand 
have explored proofs by mathematical induction of hardware and software 
systems with some limitations, e.g., requiring changes in the code 
to introduce loop invariants~\cite{Donaldson10,GrosseLD09}.~This 
complicates the automation of the verification process,
unless other methods are used in combination to automatically compute the loop 
invariant~\cite{S_sharma:2011,S_ancourt:2010}. 
Similar to the approach proposed by~\cite{HagTin:2008,Kinductor}, 
our method is completely automatic and does not require the user to provide 
loops invariants as the final assertions after each loop.
On the other hand, state-of-the-art BMC tools have been widely used, 
but as bug-finding tools since they typically analyze 
bounded program runs~\cite{CBMC2012,MerzFS12}.~This paper closes this gap, 
providing clear evidence that the \textit{k}-induction algorithm in combination with invariants 
can be applied to a broader range of C programs without manual intervention.

Gro{\ss}e et al. describe a method to prove properties of Transaction Level Modeling designs in SystemC~\cite{GrosseLD09}. 
The approach consists of converting a SystemC program into a C program, 
and then it performs the proof of the properties by mathematical induction 
using the CBMC tool~\cite{CBMC2012}. 
The difference to the one described in this paper lies on the transformations carried out in the forward condition. 
During the forward condition, 
transformations similar to those inserted during the inductive step 
in our approach, are introduced in the code to check whether 
there is a path between an initial state and the current state $k$; 
while the algorithm proposed in this paper, an assertion is inserted 
at the end of the loop to verify that all states are reached in $k$ steps. 

Donaldson et~al.\ describe a verification tool called Scratch~\cite{Donaldson10} 
to detect data races during Direct Memory Access in the CELL BE processor 
from IBM~\cite{Donaldson10}. The approach used to verify C programs is 
\textit{k}-induction, which is implemented in the Scratch tool using two steps: 
the base case and the inductive step. Scratch can prove the absence of data races, but it is 
restricted to verify that specific class of problems for a particular type of hardware. 
The steps of the algorithm are similar to the one proposed in this paper, but it requires annotations in the code
to introduce loops invariants. 

Kahsai et al. describe PKIND, a parallel version of the tool KIND, 
used to verify invariant properties of programs written in Lustre~\cite{Kahsai11}. 
To verify a Lustre program, PKIND starts three processes, 
one for base case, one for inductive step, and one for invariant generation, 
note that unlike ESBMC, the k-induction algorithm used by PKIND does not have a forward 
condition step. 
This happens because PKIND is used for Lustre programs that do not terminate. 
Hence, there is no need for checking whether loops have been unrolled completely.  
The base case starts the verification with $k=0$, and increments its value 
until it finds a counterexample or it receives a message from the inductive step process that 
a solution was found. 
Similarly, the inductive step increases the value 
of $k$ until it receives a message from the base case process or a solution is found. 
The invariant generation process generates a set of candidates invariants from predefined 
templates and constantly feeds the inductive step process, as done recently by 
Beyer et al.~\cite{Beyer:2015arxiv}. 

% -----------------------------------------------------------------
% CONCLUSIONS AND FUTURE WORKS - 5 -> [DOING]
% -----------------------------------------------------------------
% 
%=-=-=-=-=-=-=-=-=-=-=-=-=-=-=-=-=-=-=-=-=-=-=-=-=-=-=
\section{Conclusions}
\label{sec:conc}
%=-=-=-=-=-=-=-=-=-=-=-=-=-=-=-=-=-=-=-=-=-=-=-=-=-=-=

The main contributions of this work are the design, implementation, and evaluation of the 
\textit{k}-induction algorithm based on invariants, as well as, the use of the technique 
for the automated verification of reachability properties in embedded systems programs. 
To the best of our knowledge, this paper marks the first application of the \textit{k}-induction algorithm 
to a broader range of embedded C programs.
To validate the \textit{k}-induction algorithm, experiments were performed 
involving $142$ benchmarks of the SV-COMP 2015 \textit{loops} subcategory, and 
$34$ ANSI-C programs from the embedded systems benchmarks. 
Additionally, we presented a comparison to the ESBMC with $k$-induction, 
CBMC with $k$-induction, and CPAChecker with $k$-induction and invariants.

The experimental results are promising; the proposed method adopting 
\textit{k}-induction with invariants (DepthK) determined $11.27$\% more accurate results than that obtained by 
CPAChecker, which had the second best result in the SV-COMP $2015$ 
loops subcategory. 
The experimental results also show that the \textit{k}-induction algorithm without invariants 
was able to verify $49.29\%$ of the programs in the SV-COMP benchmarks, 
% in $141.58$ min, 
and \textit{k}-induction with invariants (DepthK) was able 
to verify $66.19\%$ of the benchmarks. 
% in $190.38$ min. 
We identified that \textit{k}-induction with invariants 
determined $17$\% more accurate results than the \textit{k}-induction algorithm 
without invariants.

For embedded systems benchmarks, we identified some improvements in the 
DepthK tool, related to defects in the tool execution, and possible adjustments to 
invariant generation with PIPS. This is because the results were inferior compared 
to the other tools for the embedded systems benchmarks, where DepthK only obtained better 
results than CBMC tool. However, we argued that the proposed method, in comparison to other 
state of the art tools, showed promising results indicating its effectiveness.
As future work, we will improve the robustness of DepthK and tune the PIPS parameters to produce stronger invariants.



%\bibliographystyle{IEEEtran}
% BIBLIOGRAPHY
% -----------------------------------------------------------------
\vspace{-1ex}
\bibliographystyle{IEEEtran}
\bibliography{references}

\end{document}